\documentclass[10pt,twocolumn,letterpaper]{article}

\usepackage{fancyhdr}

\usepackage{cvpr}
\usepackage{times}
\usepackage{epsfig}
\usepackage{graphicx}
\usepackage{amsmath}
\usepackage{amssymb}

\usepackage[pagebackref=true,breaklinks=true,letterpaper=true,colorlinks,bookmarks=false]{hyperref}

\newcommand{\argmax}{\operatornamewithlimits{argmax}}

\newcommand{\myparatight}[1]{\smallskip\noindent{\bf {#1}:}~}

\cvprfinalcopy 

\graphicspath{ {../fig/} }

\begin{document}

\title{On Certifying Robustness against Backdoor Attacks via Randomized Smoothing}


\author{Binghui Wang, Xiaoyu Cao, Jinyuan Jia, Neil Zhenqiang Gong\\
Duke University\\
{\tt\small \{binghui.wang,xiaoyu.cao,jinyuan.jia,neil.gong\}@duke.edu}
}


\maketitle

\thispagestyle{fancy}


\begin{abstract}
\label{abstract}
Backdoor attack is a severe security threat to deep neural networks (DNNs). 
We envision that, like adversarial examples, there will be a cat-and-mouse game for backdoor attacks, i.e., new empirical defenses are developed to defend against backdoor attacks but they are soon broken by strong  adaptive backdoor attacks. To prevent such cat-and-mouse game, we take the first step towards certified defenses against backdoor attacks. Specifically, in this work, we study the feasibility and effectiveness of certifying robustness against backdoor attacks using a recent technique called randomized smoothing. Randomized smoothing was originally developed to certify robustness against adversarial examples. We generalize randomized smoothing to defend against backdoor attacks. Our results show the theoretical feasibility of using randomized smoothing to certify robustness against backdoor attacks. However, we also find that existing randomized smoothing methods have limited effectiveness at defending against backdoor attacks, which highlight the needs of new theory and methods to certify robustness against backdoor attacks.
\end{abstract}

\section{Introduction}
\label{intro}
{Backdoor attack}~\cite{chen2017targeted,liu2018trojaning,gu2017badnets,yao2019latent} is a severe security threat to  DNNs. 
Specifically, in a backdoor attack, an attacker adds a trigger to the features of some training examples and changes their labels to a \emph{target label} during the training or fine-tuning process. Then, when the attacker adds the same trigger to the features of a testing example, the learnt classifier predicts the target label for the testing example with
the trigger. We envision that, like adversarial examples, there will be a cat-and-mouse game for backdoor attacks. Indeed, various empirical defenses~\cite{chen2018detecting,chou2018sentinet,gao2019strip,chen2019deepinspect,qiao2019defending,wang2019neural,liu2019abs} have been proposed to defend against backdoor attacks in the past few years. For instance, Neural Cleanse~\cite{wang2019neural} aims to detect and reconstruct the trigger via solving an optimization problem. However,  \cite{guo2019tabor} showed that Neural Cleanse  fails to detect the trigger when the trigger has different sizes, shapes, and/or locations. 

To prevent such cat-and-mouse game, we take the first step towards certifying robustness against backdoor attacks. 
Specifically, we study the feasibility and effectiveness of certifying robustness against backdoor attacks using \emph{randomized smoothing}~\cite{cao2017mitigating,cohen2019certified,lee2019stratified,jia2020WWWcertified}. 
Randomized smoothing was originally developed to certify robustness against adversarial examples~\cite{Szegedy14,goodfellow2014explaining,carlini2017towards}. In particular, Cao and Gong~\cite{cao2017mitigating} is the first to propose randomized smoothing as an empirical defense against adversarial examples (they call it \emph{region-based classification}). Cohen et al.~\cite{cohen2019certified} derived a tight certified robustness guarantee for randomized smoothing with Gaussian noise using the Neyman-Pearson Lemma~\cite{neyman1933ix}.  
Lee et al.~\cite{lee2019stratified} and Jia et al.~\cite{jia2020WWWcertified} generalized randomized smoothing to discrete data using discrete noise distribution. 

In this work, we generalize randomized smoothing to defend against backdoor attacks. Given  an arbitrary function (we call it \emph{base function}), which takes a data vector as an input and outputs a label, randomized smoothing can turn the function to be a provably robust one via adding random noise to the input data vector. Specifically, a function is provably robust if it outputs the same label for all data points in a region (e.g., $\ell_p$-norm ball) around an input. Our idea to certify robustness against backdoor attacks consists of two steps. First, we view the entire process of learning a classifier from the training dataset and using the classifier to make predictions for a testing example as a base function. In particular, this base function takes a training dataset and a testing example as an input, and the base function outputs a predicted label for the testing example. Second, we add random  noise to the training dataset and a testing example to overwhelm the trigger that the attacker injects to the training examples and the testing example.

We evaluate our method on a subset of the MNIST dataset.  
Our defense guarantees that 36\% of testing images can be classified correctly when an attacker arbitrarily perturbs at most 2 pixels/labels of the training examples and pixels of a testing example.  
Our results show the theoretical feasibility of using randomized smoothing to certify robustness against backdoor attacks. 
However, our results also show that existing randomized smoothing methods with additive noise have limited  effectiveness at defending against backdoor attacks. Our study highlights the needs of new theory and techniques to certify robustness against backdoor attacks.

\section{Background on Randomized Smoothing}
\label{background}

Randomized smoothing is state-of-the-art technique to certify the robustness of a classifier against adversarial examples. Randomized smoothing~\cite{cao2017mitigating,liu2018towards} was first proposed as an empirical defense against adversarial examples. For instance, Cao and Gong~\cite{cao2017mitigating} proposed to add uniform random noise from a hypercube centered at a testing example and use majority vote to smooth the predicted label of the testing example. They called the method \emph{region-based classification} as it leverages information in a region around a testing example to predict its label. Lecuyer et al. ~\cite{lecuyer2018certified} derived the first certified robustness guarantee for randomized smoothing using differential privacy techniques. Li et al.~\cite{li2018second} derived a tighter certified robustness guarantee using information-theoretic techniques. Cohen et al.~\cite{cohen2019certified} derived a tight certified robustness guarantee for randomized smoothing with Gaussian noise using the Neyman-Pearson Lemma~\cite{neyman1933ix}. Jia et al.~\cite{jia2020certified} derived a tight certified robustness guarantee of general top-$k$ predictions for randomized smoothing with Gaussian noise. Lee et al.~\cite{lee2019stratified} and Jia et al.~\cite{jia2020WWWcertified} generalized randomized smoothing to discrete data using discrete noise. We will use such randomized smoothing for discrete data because labels are discrete/categorical. Salman et al.~\cite{salman2019provably} and Zhai et al.~\cite{zhai2020macer} proposed methods to train classifiers that have better certified robustness under randomized smoothing.

Next, we describe randomized smoothing from a general function perspective, making it easier to understand how we apply randomized smoothing to certify robustness against backdoor attacks. 

\subsection{Building a Smoothed Function}
\label{RS_discrete}

\myparatight{Data vector $v$} 
Suppose we have a data vector $v$. We consider each dimension of $v$ to be discrete, as many applications have discrete data, e.g., pixel values are discrete. Moreover, when certifying robustness against backdoor attacks, some dimensions of $v$ correspond to the labels of the training examples, which are discrete.   
Without loss of generality, we assume each dimension of $v$ is from the discrete domain $\{0,\frac{1}{d}, \cdots, \frac{d-1}{d}\}$, where $d$ is the domain size. 
We note that randomized smoothing could also be applied when the dimensions of $v$ have different domain sizes. However, for simplicity, we assume the dimensions have the same domain size. 

\myparatight{Base function} Suppose we have
an arbitrary function (we call it \emph{base function}), which takes the data vector $v$ as an input and outputs a label in a set $\{0,1,\cdots,c-1\}$. 
For instance, when certifying robustness against adversarial examples, the base function is a classifier whose robustness we aim to certify. 
When certifying robustness against backdoor attacks, we treat the entire process of learning a classifier and using the learnt classifier to make predictions for a testing example as a base function. For convenience, we denote the base function as $f$, and $f(v)$ is the predicted label for $v$. 

\myparatight{Adversarial perturbation} An attacker can perturb the data vector $v$. We denote by $\delta$ the adversarial perturbation an attacker adds to the vector $v$, where $\delta_j$ is the perturbation added to the $j$th dimension of the vector $v$ and $\delta_j \in \{0,\frac{1}{d}, \cdots, \frac{d-1}{d}\}$.  Moreover, we denote by ${v} \oplus \delta$ the \emph{perturbed data vector}, where the operator  $\oplus$ is defined for each dimension as follows:
{
\begin{align}
\label{noisydata}
v_j \oplus \delta_j = \frac{(v_j\cdot d + \delta_j\cdot d) \text{ mod } d}{d},
\end{align}
}%
where $v_j$ and $\delta_j$ are the $j$th dimensions of the vector $v$ and  the perturbation vector $\delta$, respectively. We measure the magnitude of the adversarial perturbation using its $\ell_0$ norm, i.e., $||\delta||_0$. We adopt $\ell_0$ norm because it is semantically easy to interpret. In particular, $\ell_0$ norm of the adversarial perturbation is the number of dimensions of $v$ that are arbitrarily perturbed by the attacker.

\myparatight{Smoothed function}  Randomized smoothing builds a new function from the base function via adding random noise to the data vector $v$. We call the new function \emph{smoothed function}. Specifically, we denote by $\epsilon$ the random noise vector, where the $j$th dimension $\epsilon_j\in \{0,\frac{1}{d}, \cdots, \frac{d-1}{d}\}$ is the random noise added to $v_j$. We consider $\epsilon_j$ has the following distribution~\cite{lee2019stratified,jia2020WWWcertified}: 
{
\begin{equation}
\label{discretenoisedistribution}
\begin{aligned}
& \text{Pr}(\epsilon_j=0)=\beta, \\
& \text{Pr}(\epsilon_j=t)=\theta=\frac{1-\beta}{d-1}, \, \forall t\in \{\frac{1}{d}, \frac{2}{d}, \cdots, \frac{d-1}{d}\}.
\end{aligned}
\end{equation}
}%
Moreover, $v\oplus \epsilon$ is the \emph{noisy data vector}, where the operator $\oplus$ is defined in Equation~\ref{noisydata}. The noise distribution indicates that, when adding a random noise vector $\epsilon$ to the data vector $v$, the $j$th dimension of $v$ is preserved with a probability $\beta$ and is changed to any other value with a probability $\frac{1-\beta}{d-1}$. 
Since we add random noise to the data vector $v$, the base function $f$ outputs a random label. We define a smoothed function $g$, which outputs the label with the largest probability as follows:
{
\begin{equation}
g(v) = \argmax_{y \in \{0,1,\cdots, c-1 \}} \text{Pr}(f(v \oplus \epsilon) = y), 
\end{equation}
}%
where $g(v)$ is the label predicted for $v$ by the smoothed function. Note that $g(v\oplus \delta)$ is the label predicted for the perturbed data vector $v\oplus \delta$.

\subsection{Computing Certified Radius}
 Randomized smoothing guarantees that the smoothed function $g$ predicts the same label when the adversarial perturbation $\delta$ is bounded. In particular, according to~\cite{lee2019stratified}, we have:
 {
 \begin{align}
 \label{topkradius}
  g({v} \oplus \delta)=l,\ \forall ||\delta||_0 \leq R(\underline{p_l}),
  \end{align}
  }%
where $\underline{p_l}\leq \text{Pr}(f(v \oplus \epsilon) =l)$ is a lower bound of the probability $\text{Pr}(f(v \oplus \epsilon)=l)$ that $f$ predicts a label $l$ when adding random noise $\epsilon$ to $v$, and $R(\underline{p_l})$ is called \emph{certified radius}. Intuitively, Equation~\ref{topkradius} shows that the smoothed function $g$ predicts the same label $l$ when an attacker arbitrarily perturbs at most $R(\underline{p_l})$ dimensions of the data vector $v$. 
Note that the certified radius $R(\underline{p_l})$ depends on $\underline{p_l}$. In particular, given any lower bound $\underline{p_l}$, we can compute the certified radius $R(\underline{p_l})$.  The computation details can be found in \cite{lee2019stratified}.  Estimating a lower bound  $\underline{p_l}$ is the key to compute the certified radius. Next, we describe how to estimate $\underline{p_l}$.

Cohen et al.~\cite{cohen2019certified} proposed a Monte Carlo method to predict $l$ and estimate $\underline{p_l}$ with probabilistic guarantees. Specifically, we sample $N$ noise $\epsilon^1, \epsilon^2, \cdots, \epsilon^N$ from the noise distribution defined in Equation~\ref{discretenoisedistribution}. We compute the label $f(v \oplus \epsilon^j)$ for each noise $\epsilon^j$, and we compute the label frequency $count[i]=\sum_{j=1}^N \mathbb{I} (f(v \oplus \epsilon^j)=i)$ for each $i \in \{0,1,\cdots, c-1\}$, where $\mathbb{I}$ is an indicator function. The smoothed function $g$ outputs the label $l$ with the largest frequency. Moreover, Cohen et al.~\cite{cohen2019certified} proposed to use the Clopper-Pearson method~\cite{brown2001interval}  to estimate $\underline{p_l}$ as follows:
{
\begin{align}
\label{cp}
 \underline{p_l}=B(\alpha; count[l], N-count[l]+1), 
 \end{align}
 }%
where $1-\alpha$ is the confidence level and $B(\alpha; a,b)$ is the $\alpha$th quantile of the Beta distribution with shape parameters $a$ and $b$.

\section{Certifying Robustness against Backdoor Attacks}
\label{robustness}

Suppose we have a training dataset  $\{\mathbf{X}, \mathbf{y}\}=\{(x_1, y_1), (x_2,y_2),\cdots,(x_T,y_T)\}$, where $x_i$ and $y_i$ are the feature vector and label of the $i$th training example, respectively. 
Suppose further we have a  learning algorithm $\mathcal{A}$ which takes the training dataset as an input and produces a classifier $h$, i.e., $h=\mathcal{A}(\mathbf{X}, \mathbf{y})$.  
We use the classifier $h$ to predict the label for a testing example $x$.  
We generalize randomized smoothing to certify robustness against backdoor attacks. 
Our key idea is to  combine the entire process of training and prediction as a single function $f(\mathbf{X}, \mathbf{y}, x)$, which is the predicted label for a testing example $x$ when the classifier is trained on $\{\mathbf{X}, \mathbf{y}\}$ using the algorithm $\mathcal{A}$. We view the function $f$ as a base function and apply randomized smoothing to it. 

\myparatight{Constructing a smoothed function} 
We view the concatenation of the feature matrix $\mathbf{X}$, the label vector $\mathbf{y}$, and the features of the testing example $x$ as the data vector $v$ that we described in Section~\ref{background}.
We add a random noise matrix $\tau$ to the feature matrix $\mathbf{X}$, where each entry of the noise matrix is drawn from the  distribution defined in Equation~\ref{discretenoisedistribution} with $d$ as the feature domain size.
We add a random noise vector $\epsilon$ to the label vector  $\mathbf{y}$, where each entry of the noise vector is drawn from the  distribution defined in Equation~\ref{discretenoisedistribution} with $d=c$. 
Furthermore, we add a random noise vector $\gamma$ to the testing example $x$, where each entry of the noise vector is drawn from the  distribution defined in Equation~\ref{discretenoisedistribution} with $d$ as the feature domain size.
Since we add random noise, the output of the base function $f$ is also random. The smoothed function $g$ outputs the label that has the largest probability.  
Formally, we have:
\begin{align}
g(\mathbf{X}, \mathbf{y}, x)=\argmax_{y \in \{0,1,\cdots, c-1 \}} \text{Pr}(f(\mathbf{X} \oplus \tau, \mathbf{y}\oplus \epsilon, x \oplus \gamma)=y), 
\end{align}
where $g(\mathbf{X}, \mathbf{y}, x)$ is the label predicted by the smoothed function for $x$. 

\myparatight{Computing the certified radius}
We denote by a matrix $\delta_1$ and a vector $\delta_2$ the perturbations an attacker adds to the feature matrix $\mathbf{X}$ and label vector $\mathbf{y}$, respectively. Moreover, we denote by a vector $\delta_3$ the adversarial perturbation an attacker adds to the testing example $x$. 
Based on Equation~\ref{topkradius}, we have the following:
{
\begin{equation}
\label{soln_trojan}
\begin{aligned}
& g(\mathbf{X} \oplus \delta_1, \mathbf{y} \oplus \delta_2, x \oplus \delta_3)=l, \\
& \forall ||\delta_1||_0 +||\delta_2||_0+||\delta_3||_0 \leq R(\underline{p_l}),
\end{aligned}
\end{equation}
}%
where $\underline{p_l}\leq \text{Pr}(f(\mathbf{X} \oplus \tau, \mathbf{y}\oplus \epsilon, x \oplus \gamma)=l)$. 
Equation~\ref{soln_trojan} means that the smoothed function $g$ predicts the same label for the testing example $x$ when the $\ell_0$ norm of the adversarial perturbation added to the feature matrix and label vector of the training examples as well as the testing example is bounded by $R(\underline{p_l})$. 
The key to computing the certified radius $R(\underline{p_l})$ is to estimate $\underline{p_l}$. We use the Monte Carlo method described in Section~\ref{background} to estimate $\underline{p_l}$. Specifically, we randomly sample $N$ noise matrices $\tau^1, \tau^2, \cdots, \tau^N$, $N$ noise vectors $\epsilon^1, \epsilon^2, \cdots, \epsilon^N$, as well as $N$ noise vectors  $\gamma^1, \gamma^2, \cdots, \gamma^N$. We train $N$ classifiers, where the $j$th classifier $h_j$ is trained using the training dataset $(\mathbf{X} \oplus \tau^j, \mathbf{y} \oplus \epsilon^j)$ and learning algorithm $\mathcal{A}$. Then, we compute the frequency of each label $i$, i.e., $N_i=\sum_{j=1}^{N}\mathbb{I}(f(\mathbf{X} \oplus \tau^j, \mathbf{y} \oplus \epsilon^j, x \oplus \gamma^j)=i)$=$\sum_{j=1}^{N}\mathbb{I}(h(x \oplus \gamma^j)=i)$ for $i \in \{0,1,\cdots,c-1\}$. Finally, we can estimate $\underline{p_l}$ using Equation~\ref{cp}. 
Note that the trained $N$ classifiers can be re-used to predict the labels and compute certified radius for different testing examples. Moreover, via evenly dividing $\alpha$ among the testing examples, we can obtain a simultaneous confidence level of $1-\alpha$ for the estimated certified radius of the testing examples based on the \emph{Bonferroni Correction}. 

\begin{figure}
\center
{\includegraphics[width=0.42\textwidth]{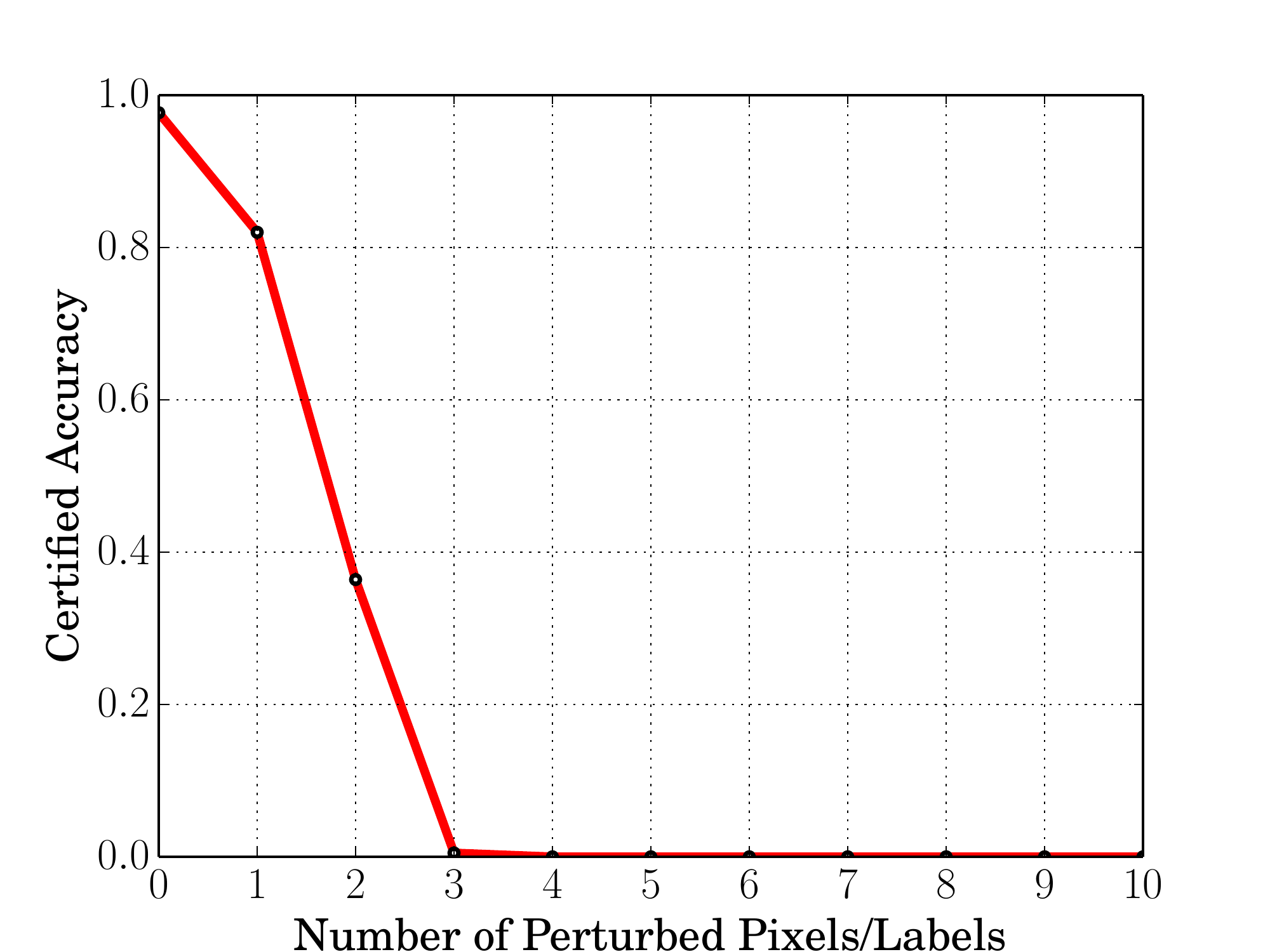}} 
\caption{Certified accuracy of our method against backdoor attacks on MNIST 1/7.}
\label{trojan_mnist}
\end{figure}

\section{Experimental Results}
\label{exp}

\myparatight{Experimental setup} We use a subset of the MNIST dataset~\cite{lecun1998gradient} which only contains the handwritten digits "1" and "7" and is used for binary classification. 
Each digit image has a size $28*28$ and has normalized pixel values within $\{0,1/255, 2/255, \cdots, 1\}$. 
For simplicity, we binarize the pixel values in our experiment.  
Specifically, if a pixel value is smaller than 0.5, we set it to be 0, otherwise we set it to be 1. 
Moreover, we randomly select 100 digits from the subset of the MNIST dataset to form the training dataset and randomly select another 1,000 digits as the testing examples. 
We use a two-layer neural network as the classifier.

\myparatight{Evaluation metric} We use \emph{certified accuracy} as an evaluation metric. Specifically, for a given number of perturbed pixels/labels, certified accuracy is the fraction of testing examples, whose  labels are correctly predicted by the smoothed function and whose certified radiuses  are no smaller than the given number of perturbed pixels/labels.

\myparatight{Experimental results}  Figure~\ref{trojan_mnist} shows the certified accuracy of our method against backdoor attacks with 
$\beta=0.9$, $N=10,000$, and $1-\alpha=99.9\%$. 
 Our method guarantees that 36\% of testing images can be classified correctly when an attacker arbitrarily perturbs at most 2  pixels/labels of the training examples and pixels of a testing example.

\section{Conclusion}
\label{conclusion}

In this work, we take the first step towards certified defenses against backdoor attacks.
In particular, we study the feasibility and effectiveness of certifying robustness against backdoor attacks 
via randomized smoothing, which was originally developed to certify robustness against adversarial examples.
Our method has two key steps. First, we treat the entire process of training a classifier and using the classifier to predict the label of a testing example as a base function. 
Second, we add noise to the training data and a testing example to overwhelm the perturbation an attacker adds in a backdoor attack. 
Our results on a subset of  MNIST demonstrate that it is theoretically feasible to certify robustness against backdoor attacks using randomized smoothing, but existing randomized smoothing methods have limited  effectiveness. Our study highlights the needs of new techniques to certify robustness against backdoor attacks.

\section*{Acknowledgements}
This work was supported by the National Science
Foundation under grant No. 1937786. Any opinions,
findings and conclusions or recommendations expressed in this
material are those of the author(s) and do not necessarily reflect
the views of the funding agencies.

{\small
\bibliographystyle{ieee_fullname}
\bibliography{refs,refs_iclr}

\begin{thebibliography}{10}\itemsep=-1pt

\bibitem{brown2001interval}
Lawrence~D Brown, T~Tony Cai, and Anirban DasGupta.
\newblock Interval estimation for a binomial proportion.
\newblock {\em Statistical science}, 2001.

\bibitem{cao2017mitigating}
Xiaoyu Cao and Neil~Zhenqiang Gong.
\newblock Mitigating evasion attacks to deep neural networks via region-based
  classification.
\newblock In {\em ACSAC}, 2017.

\bibitem{carlini2017towards}
Nicholas Carlini and David Wagner.
\newblock Towards evaluating the robustness of neural networks.
\newblock In {\em IEEE S \& P}, 2017.

\bibitem{chen2018detecting}
Bryant Chen, Wilka Carvalho, Nathalie Baracaldo, Heiko Ludwig, Benjamin
  Edwards, Taesung Lee, Ian Molloy, and Biplav Srivastava.
\newblock Detecting backdoor attacks on deep neural networks by activation
  clustering.
\newblock {\em arXiv}, 2018.

\bibitem{chen2019deepinspect}
Huili Chen, Cheng Fu, Jishen Zhao, and Farinaz Koushanfar.
\newblock Deepinspect: A black-box trojan detection and mitigation framework
  for deep neural networks.
\newblock In {\em IJCAI}, 2019.

\bibitem{chen2017targeted}
Xinyun Chen, Chang Liu, Bo Li, Kimberly Lu, and Dawn Song.
\newblock Targeted backdoor attacks on deep learning systems using data
  poisoning.
\newblock {\em arXiv}, 2017.

\bibitem{chou2018sentinet}
Edward Chou, Florian Tram{\`e}r, Giancarlo Pellegrino, and Dan Boneh.
\newblock Sentinet: Detecting physical attacks against deep learning systems.
\newblock {\em arXiv}, 2018.

\bibitem{cohen2019certified}
Jeremy~M Cohen, Elan Rosenfeld, and J~Zico Kolter.
\newblock Certified adversarial robustness via randomized smoothing.
\newblock In {\em ICML}, 2019.

\bibitem{gao2019strip}
Yansong Gao, Change Xu, Derui Wang, Shiping Chen, Damith~C Ranasinghe, and
  Surya Nepal.
\newblock Strip: A defence against trojan attacks on deep neural networks.
\newblock In {\em ACSAC}, 2019.

\bibitem{goodfellow2014explaining}
Ian~J Goodfellow, Jonathon Shlens, and Christian Szegedy.
\newblock Explaining and harnessing adversarial examples.
\newblock In {\em ICLR}, 2015.

\bibitem{gu2017badnets}
Tianyu Gu, Brendan Dolan-Gavitt, and Siddharth Garg.
\newblock Badnets: Identifying vulnerabilities in the machine learning model
  supply chain.
\newblock {\em arXiv}, 2017.

\bibitem{guo2019tabor}
Wenbo Guo, Lun Wang, Xinyu Xing, Min Du, and Dawn Song.
\newblock Tabor: A highly accurate approach to inspecting and restoring trojan
  backdoors in ai systems.
\newblock {\em arXiv}, 2019.

\bibitem{jia2020certified}
Jinyuan Jia, Xiaoyu Cao, Binghui Wang, and Neil~Zhenqiang Gong.
\newblock Certified robustness for top-k predictions against adversarial
  perturbations via randomized smoothing.
\newblock In {\em ICLR}, 2020.

\bibitem{jia2020WWWcertified}
Jinyuan Jia, Binghui Wang, Xiaoyu Cao, and Neil~Zhenqiang Gong.
\newblock Certified robustness of community detection against adversarial
  structural perturbation via randomized smoothing.
\newblock In {\em The Web Conference (WWW)}, 2020.

\bibitem{lecun1998gradient}
Yann LeCun, L{\'e}on Bottou, Yoshua Bengio, Patrick Haffner, et~al.
\newblock Gradient-based learning applied to document recognition.
\newblock {\em Proceedings of the IEEE}, 1998.

\bibitem{lecuyer2018certified}
Mathias Lecuyer, Vaggelis Atlidakis, Roxana Geambasu, Daniel Hsu, and Suman
  Jana.
\newblock Certified robustness to adversarial examples with differential
  privacy.
\newblock In {\em IEEE S \& P}, 2019.

\bibitem{lee2019stratified}
Guang-He Lee, Yang Yuan, Shiyu Chang, and Tommi~S Jaakkola.
\newblock Tight certificates of adversarial robustness for randomly smoothed
  classifiers.
\newblock In {\em NeurIPS}, 2019.

\bibitem{li2018second}
Bai Li, Changyou Chen, Wenlin Wang, and Lawrence Carin.
\newblock Second-order adversarial attack and certifiable robustness.
\newblock In {\em NeurIPS}, 2019.

\bibitem{liu2018towards}
Xuanqing Liu, Minhao Cheng, Huan Zhang, and Cho-Jui Hsieh.
\newblock Towards robust neural networks via random self-ensemble.
\newblock In {\em ECCV}, 2018.

\bibitem{liu2019abs}
Yingqi Liu, Wen-Chuan Lee, Guanhong Tao, Shiqing Ma, Yousra Aafer, and Xiangyu
  Zhang.
\newblock Abs: Scanning neural networks for back-doors by artificial brain
  stimulation.
\newblock In {\em CCS}, 2019.

\bibitem{liu2018trojaning}
Yingqi Liu, Shiqing Ma, Yousra Aafer, Wen-Chuan Lee, Juan Zhai, Weihang Wang,
  and Xiangyu Zhang.
\newblock Trojaning attack on neural networks.
\newblock In {\em NDSS}, 2018.

\bibitem{neyman1933ix}
Jerzy Neyman and Egon~Sharpe Pearson.
\newblock Ix. on the problem of the most efficient tests of statistical
  hypotheses.
\newblock {\em Philosophical Transactions of the Royal Society of London.
  Series A, Containing Papers of a Mathematical or Physical Character},
  231(694-706):289--337, 1933.

\bibitem{qiao2019defending}
Ximing Qiao, Yukun Yang, and Hai Li.
\newblock Defending neural backdoors via generative distribution modeling.
\newblock In {\em NeurIPS}, 2019.

\bibitem{salman2019provably}
Hadi Salman, Jerry Li, Ilya Razenshteyn, Pengchuan Zhang, Huan Zhang, Sebastien
  Bubeck, and Greg Yang.
\newblock Provably robust deep learning via adversarially trained smoothed
  classifiers.
\newblock In {\em NeurIPS}, 2019.

\bibitem{Szegedy14}
Christian Szegedy, Wojciech Zaremba, Ilya Sutskever, Joan Bruna, Dumitru Erhan,
  Ian Goodfellow, and Rob Fergus.
\newblock Intriguing properties of neural networks.
\newblock In {\em ICLR}, 2014.

\bibitem{wang2019neural}
Bolun Wang, Yuanshun Yao, Shawn Shan, Huiying Li, Bimal Viswanath, Haitao
  Zheng, and Ben~Y Zhao.
\newblock Neural cleanse: Identifying and mitigating backdoor attacks in neural
  networks.
\newblock In {\em IEEE S \& P}, 2019.

\bibitem{yao2019latent}
Yuanshun Yao, Huiying Li, Haitao Zheng, and Ben~Y Zhao.
\newblock Latent backdoor attacks on deep neural networks.
\newblock In {\em CCS}, 2019.

\bibitem{zhai2020macer}
Runtian Zhai, Chen Dan, Di He, Huan Zhang, Boqing Gong, Pradeep Ravikumar,
  Cho-Jui Hsieh, and Liwei Wang.
\newblock Macer: Attack-free and scalable robust training via maximizing
  certified radius.
\newblock In {\em ICLR}, 2020.

\end{thebibliography}
}

\end{document}